# Pressure-Induced Stacking Disorder and Suppression of Long-Range *Sm-type* Order in Medium-Entropy Rare-Earth Alloys

Raimundas Sereika, Matthew P. Clay, Kallol Chakrabarty, Yogesh K. Vohra

Department of Physics, University of Alabama at Birmingham, Birmingham, Alabama 35294, USA

## Abstract

Rare-earth medium-entropy alloys provide a platform for investigating how chemical disorder modifies the well-established pressure-induced structural evolution of close-packed 4*f* lanthanides. Here, we study TbHoEr and TbHoDy using synchrotron X-ray diffraction in diamond anvil cells. Both alloys transform from the ambient hexagonal close-packed (*hcp*) structure to a double hexagonal close-packed (*dhcp*) phase, while no well-resolved bulk *Sm-type* intermediate phase is observed. For TbHoEr, compression to 70 GPa further reveals a high-pressure rhombohedral *hR24* phase. Unlike the constituent heavy lanthanides, however, both alloys bypass the intermediate *Sm-type* phase. Two-dimensional diffraction images further reveal streak-like diffuse scattering in the transition region, indicating stacking disorder and limited stacking coherence along the close-packed direction. These observations indicate that the transformation proceeds through a stacking-disordered close-packed state rather than through a well-ordered bulk *Sm-type* phase. We propose that configurational disorder, local lattice distortion, stacking-fault energetics, and transformation kinetics collectively suppress the development of long-range *Sm-type* order. The results demonstrate that medium-entropy alloying can fundamentally modify pressure-induced stacking pathways in rare-earth materials under extreme conditions.

# 1. Introduction

Lanthanide (4*f*) elements and their alloys hold significant technological importance due to their large magnetic moments, high magnetic anisotropy, and exceptional magnetocaloric effects. These properties make them indispensable for advanced applications such as magnetic refrigeration, permanent magnets, laser systems, and cryogenic devices [1–3]. Notably, lanthanide-based magnetic refrigeration offers an energy-efficient alternative to conventional cooling methods. Heavy rare-earth alloys show particular promise for cryogenic applications, including hydrogen liquefaction and superconducting device cooling, where high magnetocaloric performance is critical [4,5]. Furthermore, rare-earth magnets are vital components in electric motors, wind turbines, and modern electronics[6,7], underscoring their broad utility. Therefore, understanding the structural stability of rare-earth materials under extreme conditions is essential for both fundamental research and technological applications.

In recent years, configurational entropy has emerged as an effective tool for tailoring the structural and magnetic properties of multicomponent alloys. High-entropy alloys (HEAs) and medium-entropy alloys (MEAs), comprising multiple principal elements in near-equiatomic ratios, often form chemically disordered solid solutions with properties distinct from those of their constituent elements[8–10]. Among these materials, rare-earth-based HEAs (RE-HEAs) and MEAs (RE-MEAs) are especially intriguing. Unlike conventional metallic HEAs, which typically crystallize in cubic structures, rare-earth multicomponent alloys frequently retain the hexagonal close-packed (*hcp*) structure characteristic of the constituent lanthanides phase[9,11–13]. At the same time, the coexistence of multiple rare-earth elements introduces competing magnetic moments, anisotropies, and exchange interactions, leading to complex magnetic behavior and enhanced functional properties.

The magnetocaloric performance of rare-earth entropy alloys has attracted considerable attention. For example, the quinary RE-HEA GdDyErHoTb exhibits a broad magnetocaloric response with a refrigerant capacity of approximately 627 J $kg^{-1}$ near 202 K under a magnetic-field change of 5 T[9]. More recently, Gd-rich multi-principal rare-earth alloys have demonstrated refrigerant capacities as high as 751.8 J $kg^{-1}$ near 268 K, highlighting the potential of entropy-engineered rare-earth materials for advanced cooling technologies [14]. Compared with RE-HEAs, however, RE-MEAs remain much less explored. One notable example is the equiatomic ternary alloy $Ho_{1/3}Tb_{1/3}Er_{1/3}$ (TbHoEr), which was shown to crystallize in a predominantly *hcp* structure and exhibit complex magnetic ordering[15]. Related entropy-engineered rare-earth systems have also been reported in rare-earth-site substituted intermetallic compounds, including $(Tb_{1/3}Dy_{1/3}Ho_{1/3})Ni$, further illustrating growing interest in medium-entropy rare-earth materials [16].

Despite increasing research activity on rare-earth entropy alloys, their structural response to high pressure remains largely unexplored. This is particularly important because the constituent heavy lanthanides typically adopt closely related close-packed structures whose relative stability is highly sensitive to pressure. As a result, pressure-induced phase transitions in rare-earth systems are often governed by subtle changes in stacking sequence rather than by large atomic rearrangements. Medium-entropy rare-earth alloys therefore provide a unique opportunity to investigate how configurational disorder influences the stability and evolution of close-packed structures under compression. Understanding these effects is essential for establishing whether pressure-induced transformation pathways in chemically disordered rare-earth alloys follow those of the constituent elements or exhibit distinct entropy-mediated behavior.

In this work, we investigate two equimolar ternary RE-MEAs, TbHoEr and TbHoDy, using synchrotron X-ray diffraction (XRD) in diamond anvil cells (DACs). Both alloys adopt *hcp* structures at ambient conditions and possess an ideal configurational entropy of $\Delta S_{\mathrm{config}} = R\ln 3 = 1.10R$, placing them within the medium-entropy regime. Because TbHoEr and TbHoDy differ by only one constituent element while maintaining similar crystal chemistry, they provide an excellent model system for examining how moderate configurational disorder influences pressure-induced structural transformations in close-packed rare-earth systems. By comparing their pressure-dependent phase evolution and equations of state, we aim to elucidate the role of medium-entropy alloying in governing the stability and transformation pathways of close-packed rare-earth structures under compression, and to determine whether common high-pressure behaviors emerge among ternary RE-MEAs.

## 2. Methods

### A. Sample fabrication

Nominally equimolar mixtures of high-purity elemental Tb, Ho, Er, and Dy metals (≥99.9%) were prepared according to the target compositions TbHoEr and TbHoDy. The weighed elements for each composition were cold-pressed into pellets using a hydraulic press to improve handling and contact during melting. The pellets were then placed on a water-cooled copper hearth inside a MAM-1 compact arc melter (Edmund Bühler GmbH, Bodelshausen, Germany). The chamber was evacuated and backfilled with high-purity argon to establish an inert atmosphere. Prior to melting the sample pellets, a zirconium getter was melted to reduce residual oxygen in the chamber. Each pellet was then arc-melted using a tungsten electrode. The resulting ingots were flipped and re-melted multiple times to promote chemical homogeneity. After solidification, the alloys were sectioned into smaller specimens using a diamond saw, mechanically polished, and prepared for structural and compositional characterization.

### B. Microstructural and compositional characterization

Microstructural and compositional characterization was performed at room temperature and ambient pressure. Laboratory XRD measurements were conducted using a PANalytical Empyrean diffractometer with Cu $K\alpha$ radiation to evaluate the ambient-pressure crystal structure and phase purity of the synthesized alloys. Scanning electron microscopy combined with energy-dispersive X-ray spectroscopy (SEM–EDS) was carried out using a Quanta FEG 650 microscope (FEI, Hillsboro, OR, USA) to examine surface morphology, elemental homogeneity, and possible phase partitioning. The acquired EDS maps and point/area analyses were used to assess the spatial distribution of Tb, Ho, Er, and Dy within the alloys. The uncertainty of SEM–EDS quantification was estimated to be approximately 5 at.%, typical for complex multicomponent alloys.

### C. High-pressure synchrotron X-ray diffraction measurements

High-pressure synchrotron XRD experiments were performed using the DAC technique [17–19]. Symmetric DACs equipped with flat culet diamonds of 600, 400, and 250 µm diameter were employed depending on the target pressure range. For experiments using 600 µm culets, stainless-steel gaskets were pre-indented to a thickness of approximately 80 µm and drilled to form 250 µm-diameter sample chambers. For experiments using 400 µm culets, stainless-steel gaskets were pre-indented to a thickness of approximately 40 µm and drilled to form 200 µm-diameter sample chambers. For experiments using 250 µm culets, rhenium gaskets were pre-compressed to a thickness of approximately 35 µm and drilled to form 80 µm-diameter sample chambers.

For TbHoEr, measurements were conducted under three different pressure environments: hydrostatic conditions using neon as the pressure-transmitting medium, quasi-hydrostatic conditions using silicone oil, and non-hydrostatic conditions without a pressure-transmitting medium. For TbHoDy, measurements were performed under quasi-hydrostatic conditions using silicone oil. In hydrostatic and quasi-hydrostatic experiments, ruby spheres (or a piece of copper foil) were loaded together with the sample and pressure-transmitting medium for pressure determination. In non-hydrostatic experiments, a small copper foil was loaded alongside the sample and used as an internal pressure marker. Additional details regarding high-pressure techniques and the effects of different pressure environments can be found elsewhere[17–22].

Pressure uncertainties arise from pressure-marker calibration, fitting uncertainty, pressure gradients across the sample chamber, and differences in sample loading geometry among independent DAC experiments. Taking these factors into account, the absolute pressure uncertainty is estimated to be approximately 2–3% under hydrostatic and quasi-hydrostatic conditions, with larger uncertainty possible under non-hydrostatic compression at the highest pressures. This corresponds to typical uncertainties of about ±1–2 GPa near 20 GPa and up to ±3–6 GPa near 70 GPa, depending on pressure environment and pressure gradients. Uncertainties in refined lattice parameters are generally smaller than the plotted symbols unless otherwise indicated.

High-pressure synchrotron XRD measurements were carried out at the 16-ID-B and 16-BM-D beamlines of the Advanced Photon Source, Argonne National Laboratory. Diffraction data were collected using monochromatic X-ray beams with wavelengths of $\lambda$ = 0.4246 Å (16-ID-B) and $\lambda$ = 0.4133 Å (16-BM-D) and recorded using Pilatus area detectors. The X-ray wavelength and detector geometry were calibrated using a $CeO_2$ standard. Typical exposure times were approximately 3–5 min per pattern at 16-BM-D and 1–2 s per pattern at 16-ID-B.

Two-dimensional diffraction images were azimuthally integrated into one-dimensional diffraction patterns using the DIOPTAS software package [23]. Structural analysis and refinements were performed using JANA2020[24] and GSAS-II[25]. Pressure–volume data were fitted using EoSFit[26].

## 3. Results and Discussion

### A. Ambient-pressure structural and compositional characterization

Figure 1 presents the SEM–EDS characterization of the TbHoEr alloy. The SEM image provides an overview of the polished surface and reveals no obvious secondary phases or pronounced compositional contrast. The corresponding elemental maps for Tb, Ho, and Er show homogeneous spatial distributions across the analyzed region, while the overlay image confirms uniform mixing of the three constituent elements. No evidence of elemental segregation, clustering, or phase separation is observed at the micrometer scale. Quantitative EDS analysis yields atomic concentrations of 36.3 at.% Tb, 32.4 at.% Ho, and 31.3 at.% Er, in good agreement with the nominal equiatomic composition. These results demonstrate that arc melting produced a chemically homogeneous TbHoEr medium-entropy alloy suitable for subsequent high-pressure investigations.

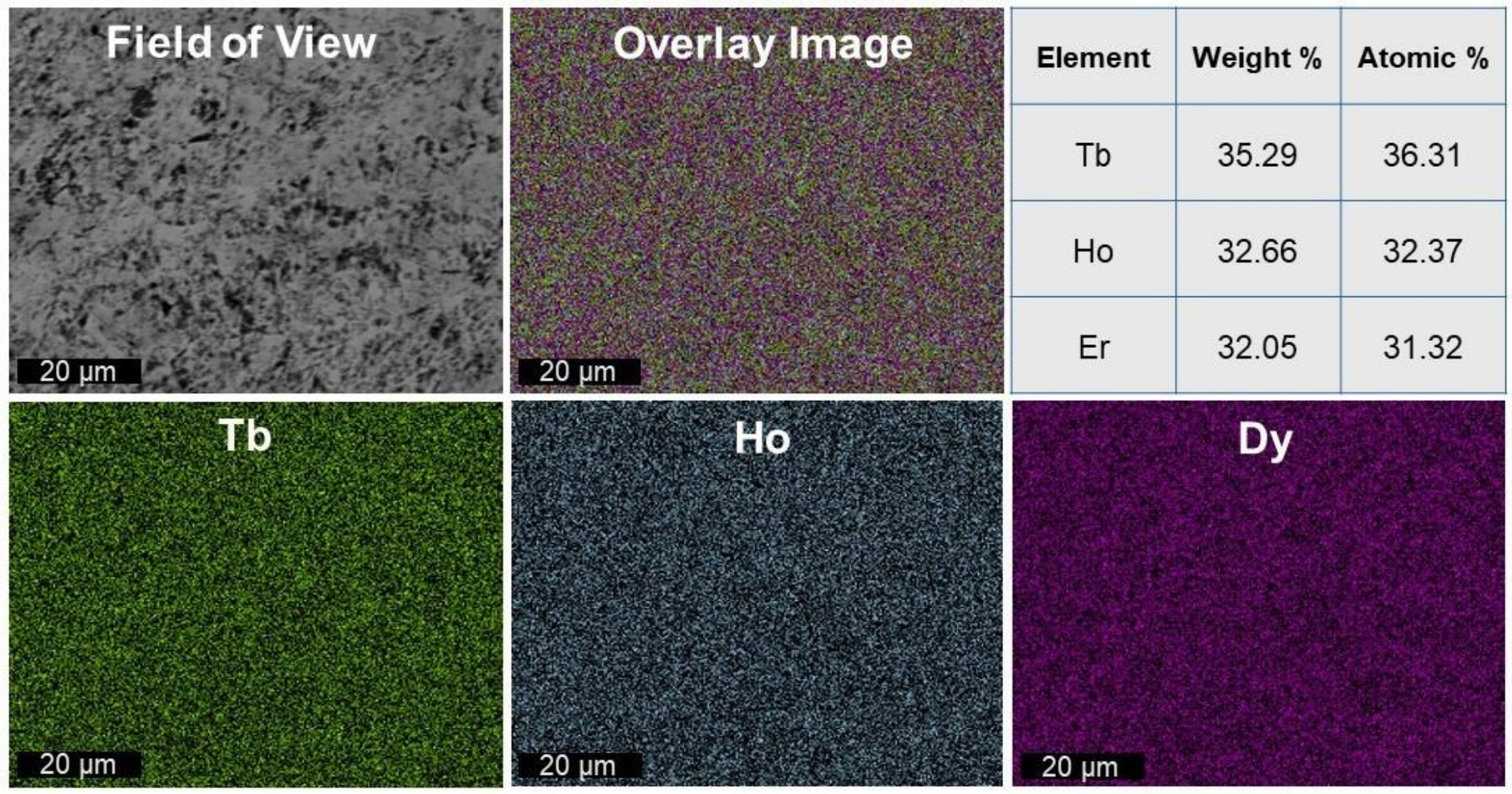

| Element | Weight % | Atomic % |
|---|---|---|
| Tb | 35.29 | 36.31 |
| Ho | 32.66 | 32.37 |
| Er | 32.05 | 31.32 |

**Figure 1.** SEM-EDS elemental distribution maps of the TbHoEr MEA. The field of view (top left) and corresponding EDS maps for Tb (green), Ho (blue), and Er (purple) display homogeneous spatial distributions, with the overlay image (top center) confirming the uniform mixing of all three elements. The quantitative EDS analysis (top right) shows near-equiatomic composition.

Figure 2 presents the SEM–EDS characterization of the TbHoDy alloy. Similar to TbHoEr, the SEM image reveals a chemically homogeneous microstructure with no obvious secondary phases or elemental segregation at the micrometer scale. The corresponding EDS elemental maps demonstrate a nearly uniform spatial distribution of Tb, Ho, and Dy throughout the analyzed region, and the overlay image confirms homogeneous mixing of the constituent elements. Quantitative EDS analysis yields atomic concentrations of 32.95 at.% Tb, 33.96 at.% Ho, and 33.09 at.% Dy, in excellent agreement with the nominal equiatomic composition. The microstructural characteristics of TbHoDy are therefore consistent with those of TbHoEr, confirming that both alloys are chemically homogeneous and near-equiatomic prior to high-pressure investigation.

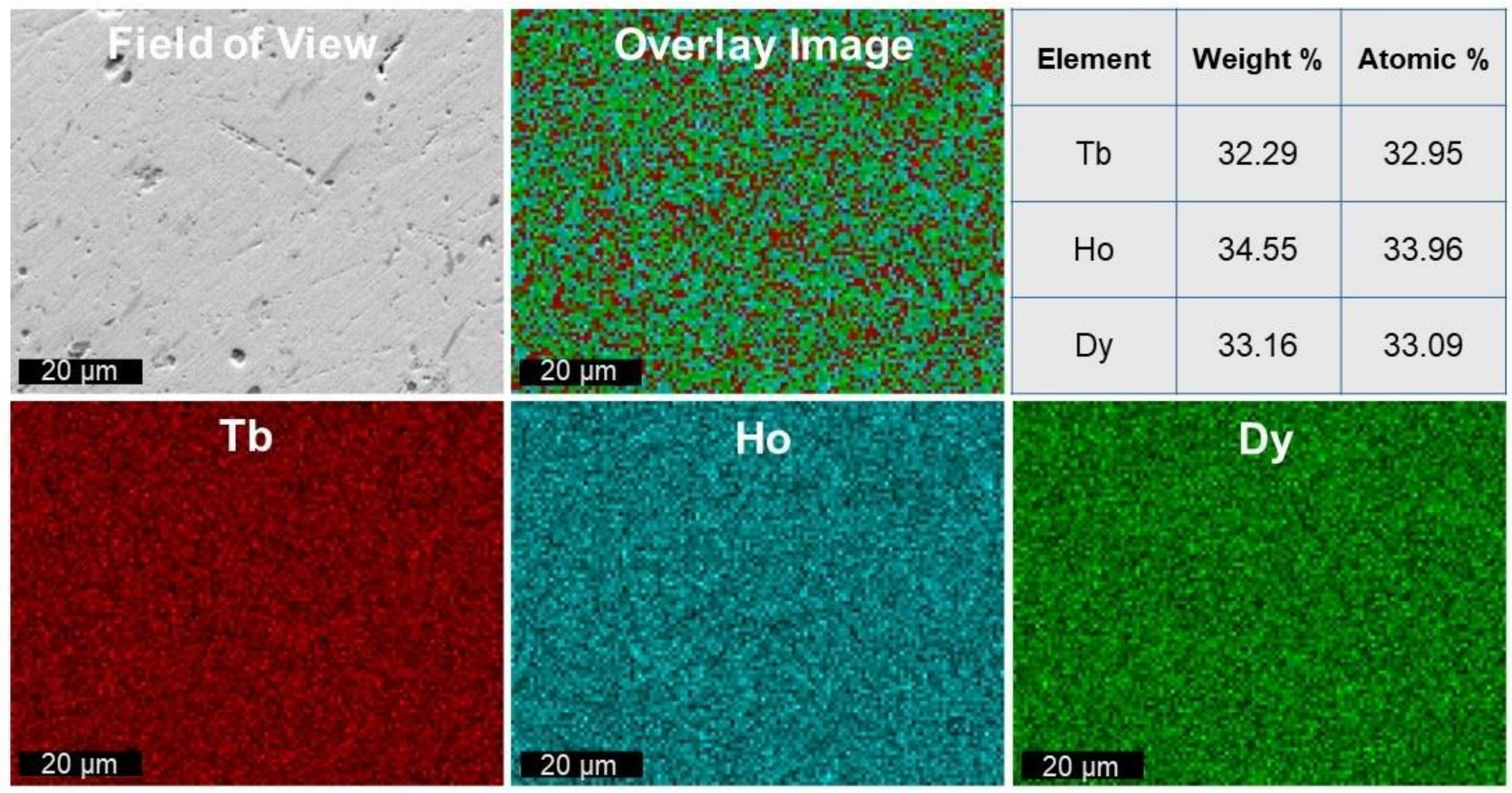

| Element | Weight % | Atomic % |
|---|---|---|
| Tb | 32.29 | 32.95 |
| Ho | 34.55 | 33.96 |
| Dy | 33.16 | 33.09 |

**Figure 2.** SEM-EDS elemental distribution maps of the TbHoDy MEA. The field of view (top left) and corresponding EDS maps for Tb (red), Ho (light blue), and Dy (green) display homogeneous spatial distributions, with the overlay image (top center) confirming the uniform mixing of all three elements. The quantitative EDS analysis (top right) shows near-equiatomic composition.

Laboratory XRD measurements further confirm that both alloys crystallize in a single-phase *hcp* structure consistent with the $P6_3/mmc$ space group (No. 194), as shown in Figs. 3 and 4. The refined lattice parameters are $a = b = 3.5835(3)$ Å and $c = 5.6362(4)$ Å for TbHoEr, and $a = b = 3.6114(4)$ Å and $c = 5.6813(1)$ Å for TbHoDy. These values are in good agreement with the composition-weighted average lattice parameters of the constituent elements (see Table S1 of the Supplemental Material), indicating that both alloys follow the expected Vegard-type behavior for substitutional rare-earth solid solutions. Notably, previously reported lattice constants for TbHoEr were substantially larger than both the elemental-average values and the present measurements, likely due to oxygen contamination or secondary phase formation[15].

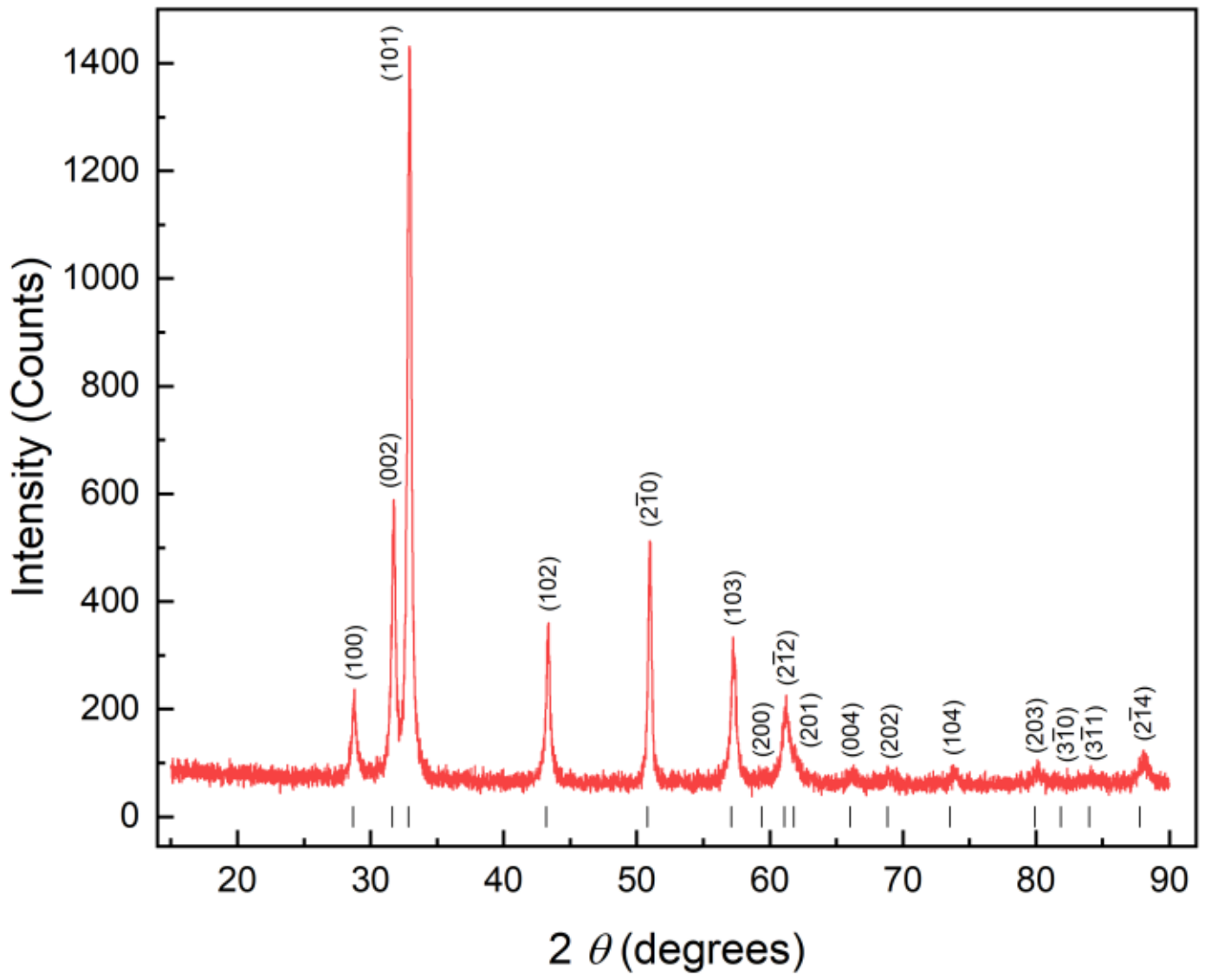

**Figure 3.** Laboratory XRD pattern of the TbHoEr medium-entropy alloy synthesized from nominally equimolar proportions of Tb, Ho, and Er. The observed reflections are indexed to the *hcp* structure with space group $P6_3/mmc$ (No. 194).

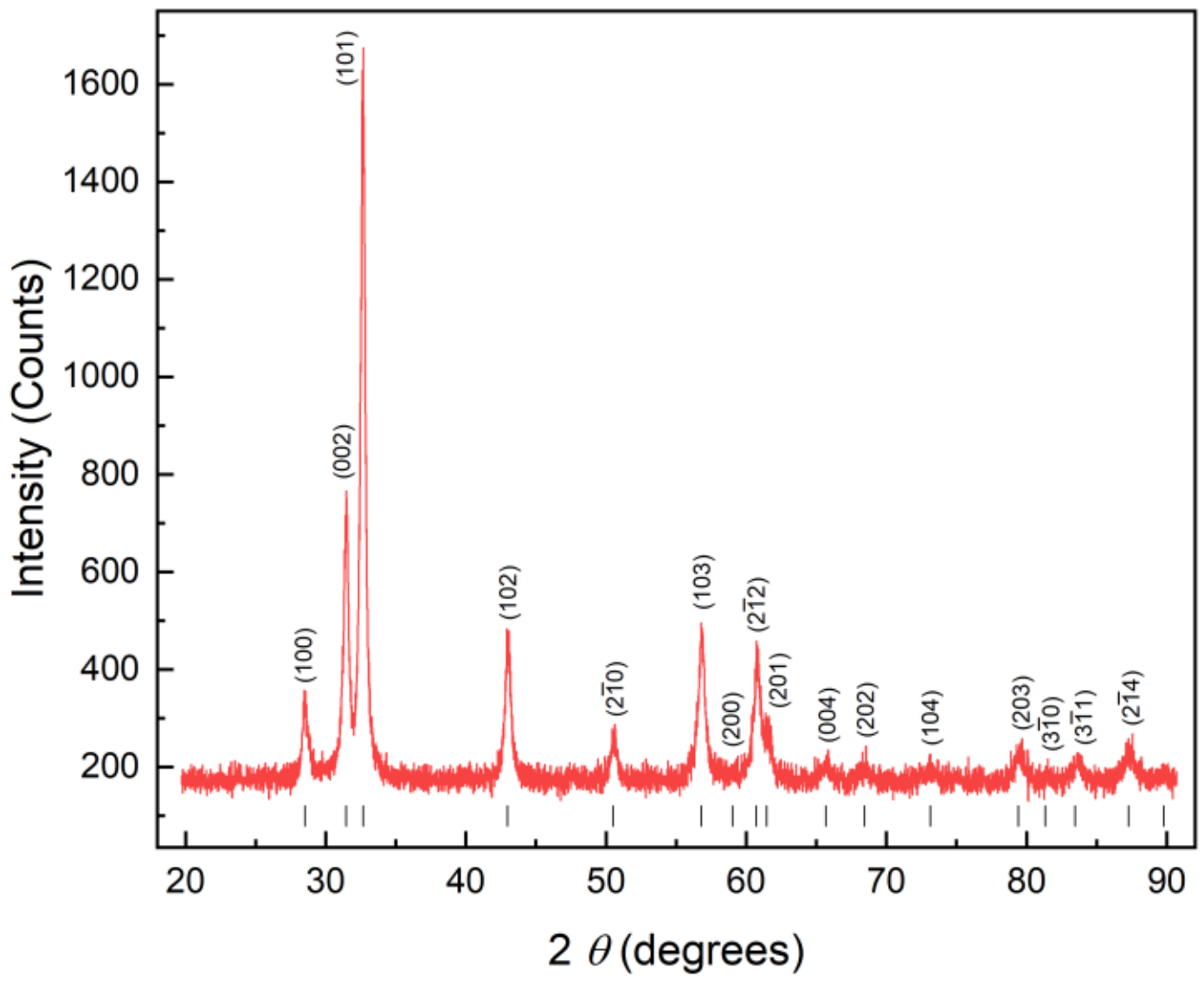

**Figure 4.** Laboratory XRD pattern of the TbHoDy medium-entropy alloy synthesized from nominally equimolar proportions of Tb, Ho, and Dy. The observed reflections are indexed to the *hcp* structure with space group $P6_3/mmc$ (No. 194).

## B. High-pressure structural evolution of TbHoEr

Synchrotron XRD measurements show that TbHoEr undergoes a pressure-induced structural transformation near 10 GPa. Figure 5 compares selected diffraction patterns collected during compression using neon and silicone-oil pressure media. Both experiments show similar behavior under hydrostatic and quasi-hydrostatic conditions, with the transformation occurring over an approximate pressure interval of 10–14 GPa. At low pressures, all observed reflections shift smoothly to higher angles with increasing pressure while retaining the ambient *hcp* indexing. Above approximately 12 GPa, clear changes appear in the diffraction profiles: additional reflections emerge, several *hcp* peaks split or change relative intensity, and the overall pattern becomes consistent with a double hexagonal close-packed structure (*dhcp*).

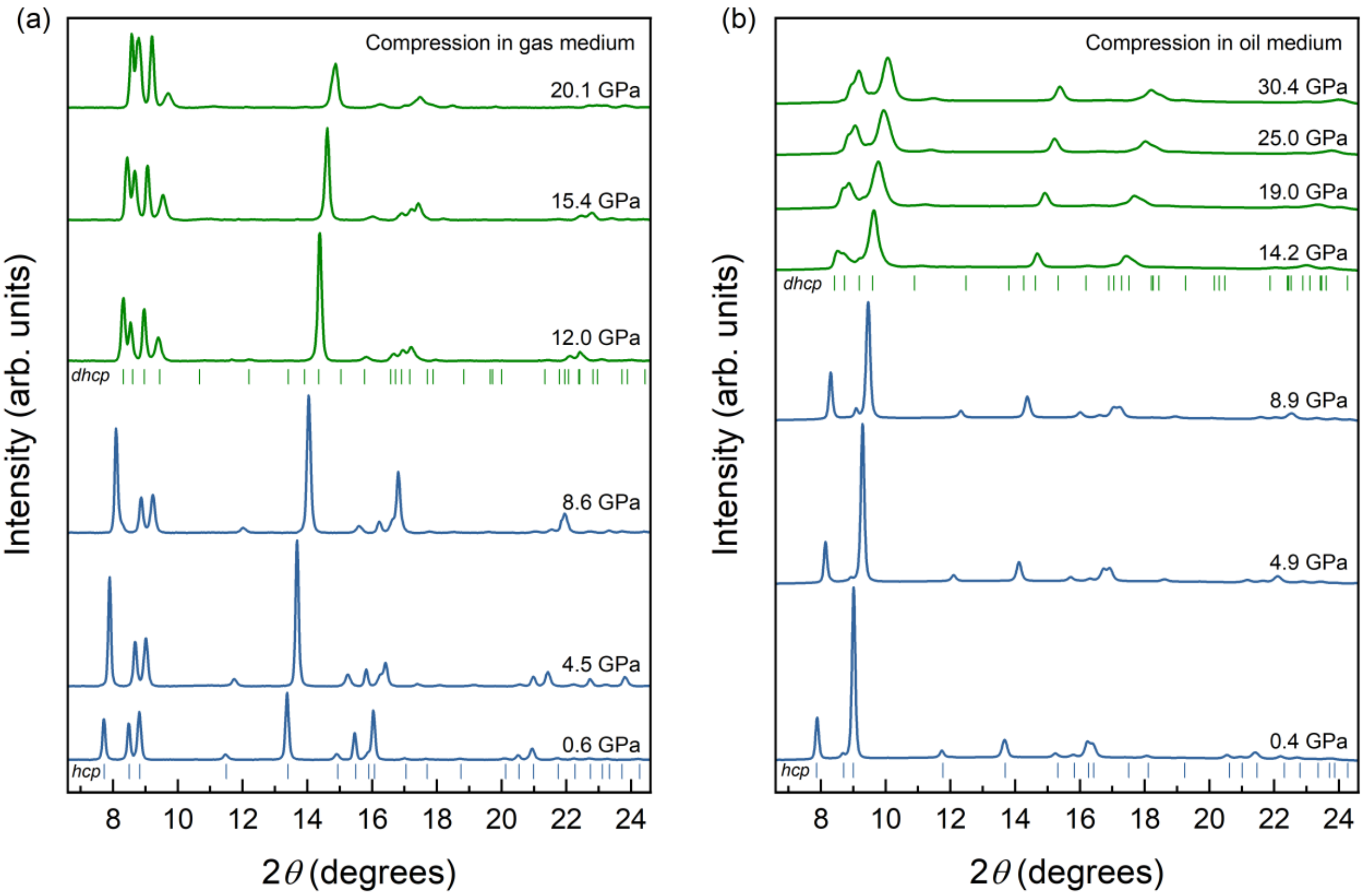

**Figure 5.** Selected synchrotron XRD patterns of TbHoEr collected during compression under (a) hydrostatic conditions using neon as the pressure-transmitting medium ($\lambda$ = 0.4133 Å) and (b) quasi-hydrostatic conditions using silicone oil ($\lambda$ = 0.4246 Å). The diffraction patterns reveal a pressure-induced structural transformation from the ambient *hcp* phase (blue) to the high-pressure *dhcp* phase (green). Vertical tick marks indicate the expected Bragg reflection positions for the corresponding structural models. Background intensity has been subtracted for clarity.

The most prominent changes occur in the low-angle region of the diffraction patterns. The strong *hcp* (100) reflection near 8.2° ($2\theta$) splits into two closely spaced peaks around 8.2 – 8.6°, which are indexed as the *dhcp* (100) and (101) reflections. The *hcp* (102) reflection also evolves into the (104) reflection, while a distinct new peak appears near 11°, corresponding to the *dhcp* (103) reflection. These features indicate a change in close-packed stacking sequence from the *hcp* ABAB arrangement to the *dhcp* ABAC arrangement, rather than a reconstructive transformation involving large atomic rearrangements.

The high-pressure diffraction patterns were examined using indexing and Le Bail refinement. Initial unconstrained indexing returned a hexagonal cell compatible with the *dhcp* repeat and suggested $P6_3/m$ (No. 176) as a possible lower-symmetry setting. However, the observed diffraction peaks do not require this reduction in symmetry. The patterns are consistently described by the conventional *dhcp* structure with ABAC stacking, for which the standard crystallographic description is $P6_3/mmc$ (No. 194). Therefore, the transformed phase is assigned to the *dhcp* structure rather than to a lower-symmetry derivative.

This structural assignment was further evaluated by comparison with the transformation pathways reported for related rare-earth systems. In previously studied lanthanide systems, including the HoDyYGdTb HEA[27] and elemental Tb, Ho, and Er under compression[28–31], pressure induces an intermediate *Sm-type* phase before the development of the *dhcp* structure. This *Sm-type* phase is often described as a 9-layer $R\bar{3}m$ structure, also known as the 9*R* or *α-Sm-type* phase. In those systems, the weakening or disappearance of characteristic *hcp* reflections is accompanied by the emergence of a systematic set of *Sm-type* reflections. In contrast, TbHoEr exhibits the loss of the *hcp* (102) reflection and the appearance of new high-pressure reflections without the characteristic diffraction pattern expected for a bulk 9*R* phase. In particular, strong

*Sm-type* reflections predicted for the $R\bar{3}m$ structure are absent, and no complete or systematic set of peaks attributable to the *Sm-type* phase is observed. Although a few weak features appear near major *hcp*/*dhcp* reflections in the transition region, they are too sparse and inconsistent to support the presence of a bulk trigonal $R\bar{3}m$ phase.

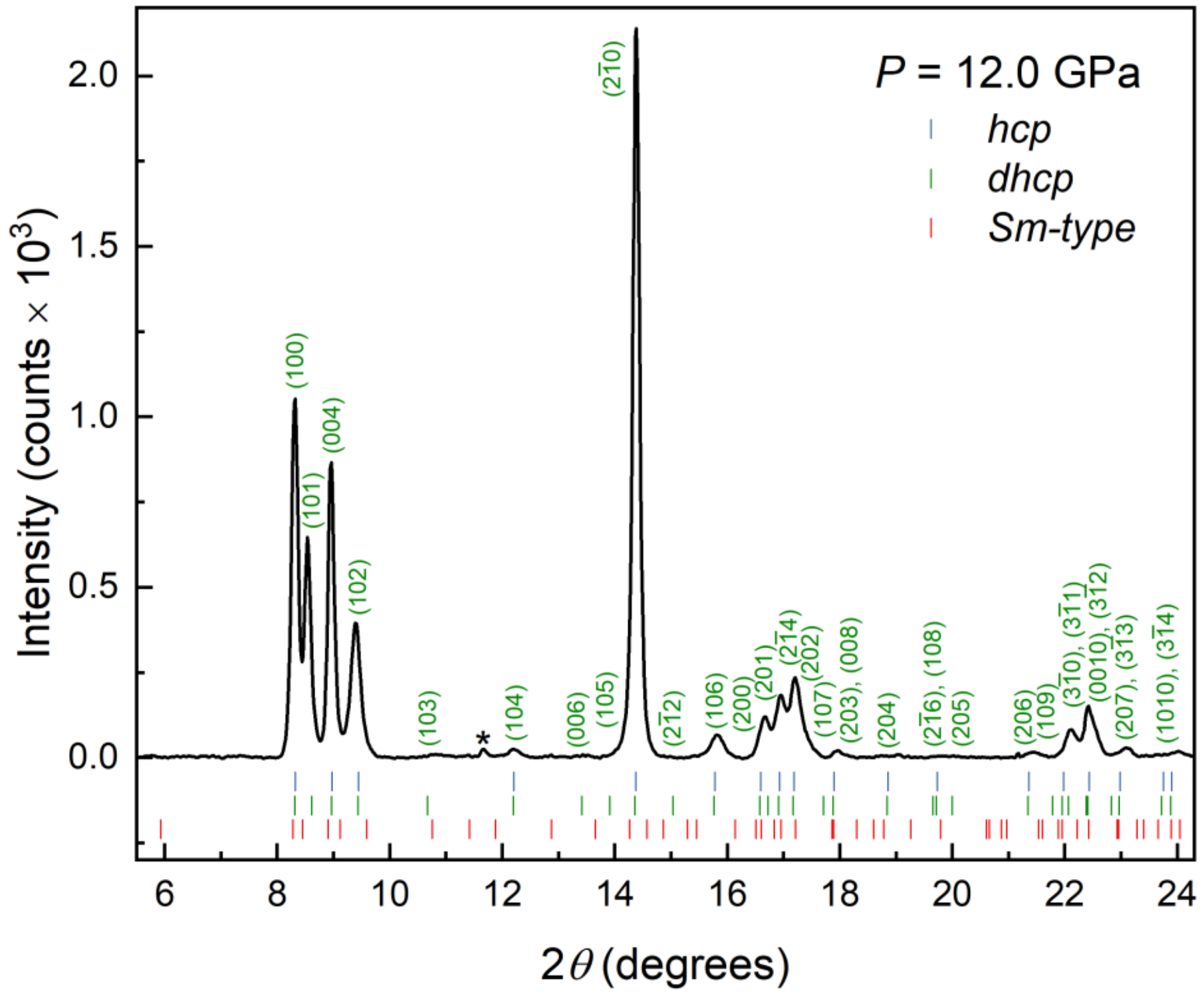

**Figure 6.** Synchrotron X-ray diffraction pattern of TbHoEr collected at 12.0 GPa using neon as the pressure-transmitting medium ($\lambda$ = 0.4133 Å). Vertical tick marks indicate calculated Bragg reflection positions for *hcp*, *dhcp*, and *Sm-type* structural models. The newly emerging reflections are indexed consistently by the *dhcp* structure, whereas characteristic *Sm-type* reflections are not observed. The asterisk marks Neon peak.

To further verify the high-pressure structure, the diffraction pattern collected at 12.0 GPa in the neon pressure medium was compared with calculated reflection positions for *hcp*, *dhcp*, and *Sm-type* structural models. As shown in Fig. 6, the newly emerging reflections are indexed consistently by the *dhcp* model. In particular, the low-angle peak splitting and the additional reflections at higher angles are accounted for by the *dhcp* structure. By contrast, the *Sm-type* model predicts several reflections that are not observed experimentally and does not provide a systematic explanation for the full set of emerging peaks. This comparison confirms that the transformation is best described as *hcp* → *dhcp*, with no evidence for a bulk intermediate *Sm-type* phase within the measured pressure range.

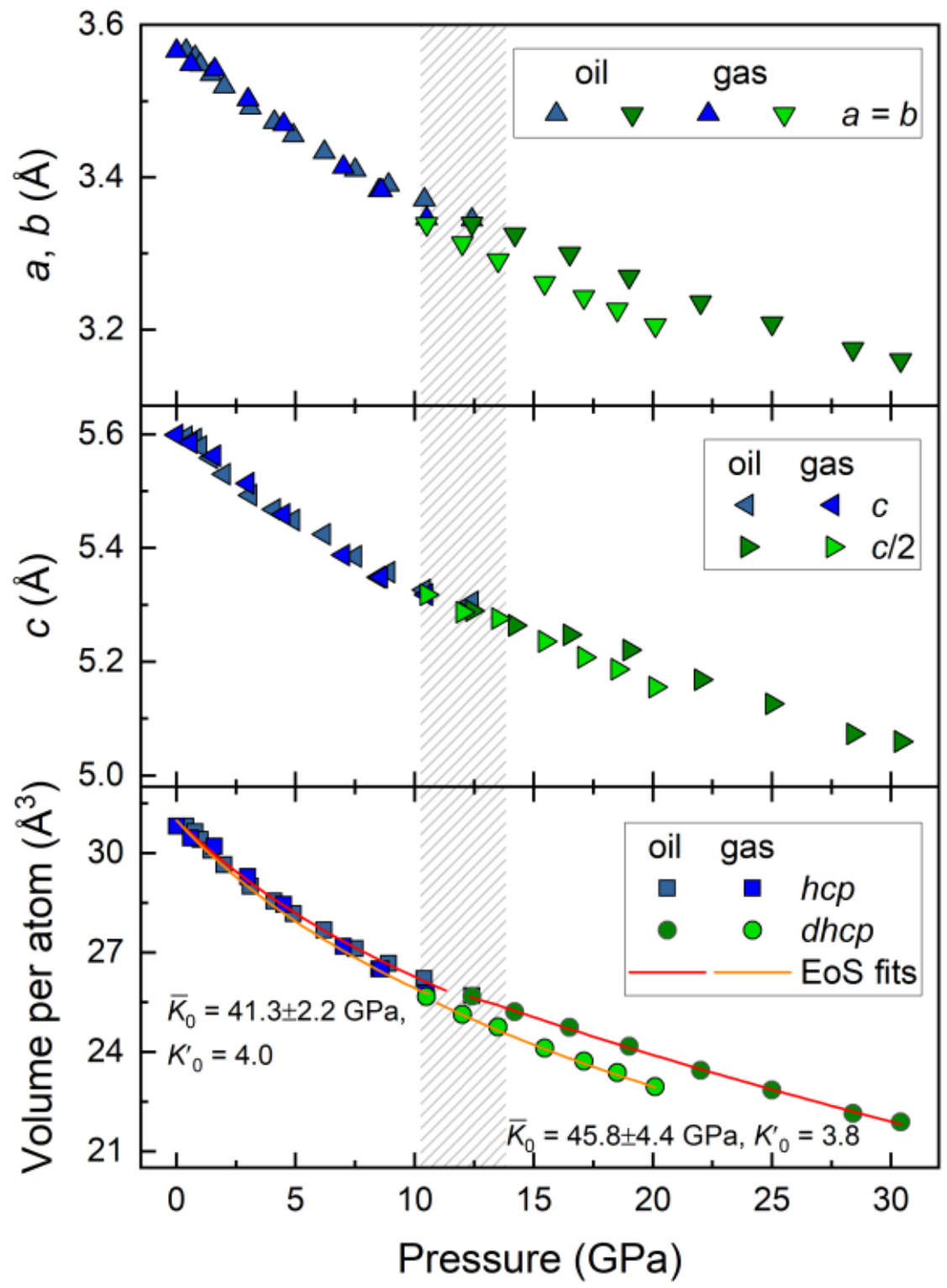

**Figure 7.** Pressure dependence of lattice parameters and atomic volume for TbHoEr extracted from compression experiments using neon and silicone-oil pressure-transmitting media. Here, symbols represent experimental data. For the *dhcp* phase, (*c*/2) is plotted to allow direct comparison with the *hcp* (*c*)-axis. The solid lines are third-order Birch–Murnaghan equation-of-state (EoS) fits to the pressure–volume data. Extracted average bulk moduli ($\overline{K}_0$) and pressure derivatives ($K_0'$) are indicated for each phase. The shaded region marks the approximate *hcp*-to-*dhcp* transition interval.

Figure 7 presents the pressure dependence of the refined lattice parameters and atomic volume for TbHoEr obtained from the neon and silicone-oil compression experiments. Because the *dhcp* unit cell contains a doubled close-packed repeat relative to *hcp*, the *dhcp* (*c*)-axis is plotted as (*c*/2) for direct comparison with the *hcp* (*c*)-axis. The halved *dhcp* (*c*)-axis smoothly continues the compression trend of the *hcp* phase across the transition region. Similarly, the ($a$)-axis and atomic volume evolve continuously through the 10–14 GPa transition interval, with no evidence of an abrupt volume collapse. The small run-to-run offsets between the neon and silicone-oil datasets are attributed primarily to pressure-calibration uncertainties, pressure gradients, and differences in DAC loading geometry, and do not affect the assigned phase sequence or the extracted average compressional trends.

The continuous evolution of lattice parameters and volume supports interpretation of the *hcp*-to-*dhcp* transformation as a stacking-sequence change from ABAB to ABAC rather than a reconstructive transformation involving major rearrangement of the metallic framework. Such behavior is consistent with pressure-induced transformations in rare-earth metals and alloys, where small energy differences between competing close-packed stacking variants can lead to gradual structural evolution over a finite pressure interval.

The pressure–volume data were fitted using a third-order Birch–Murnaghan equation of state. Averaging the results obtained from the neon and silicone-oil pressure media gives $\overline{K}_0 \approx 41.3$ GPa for the *hcp* phase and 45.8 GPa for the *dhcp* phase, with corresponding pressure derivatives of $K_0' = 4.0$ and 3.8, respectively. The pressure derivatives were fixed during fitting to stabilize the equation-of-state analysis over the limited pressure range of each phase and to allow direct comparison between the two pressure media. These values are within the range reported for compressed rare-earth metals[32], indicating that TbHoEr retains the relatively soft compressional response characteristic of lanthanide metals across the stacking transformation.

To evaluate the influence of deviatoric stress and to extend the accessible pressure range, an additional high-pressure XRD experiment was performed on TbHoEr under non-hydrostatic conditions without a pressure-transmitting medium. As expected, the absence of a pressure-transmitting medium resulted in broader and more asymmetric diffraction peaks compared with the neon and silicone-oil experiments, reflecting increased strain, pressure gradients, and preferred stress conditions within the sample chamber. These effects make detailed peak fitting and phase identification more challenging. Nevertheless, the dominant diffraction features could be consistently indexed and refined over the full compression range, allowing the pressure-dependent lattice parameters and atomic volumes to be extracted.

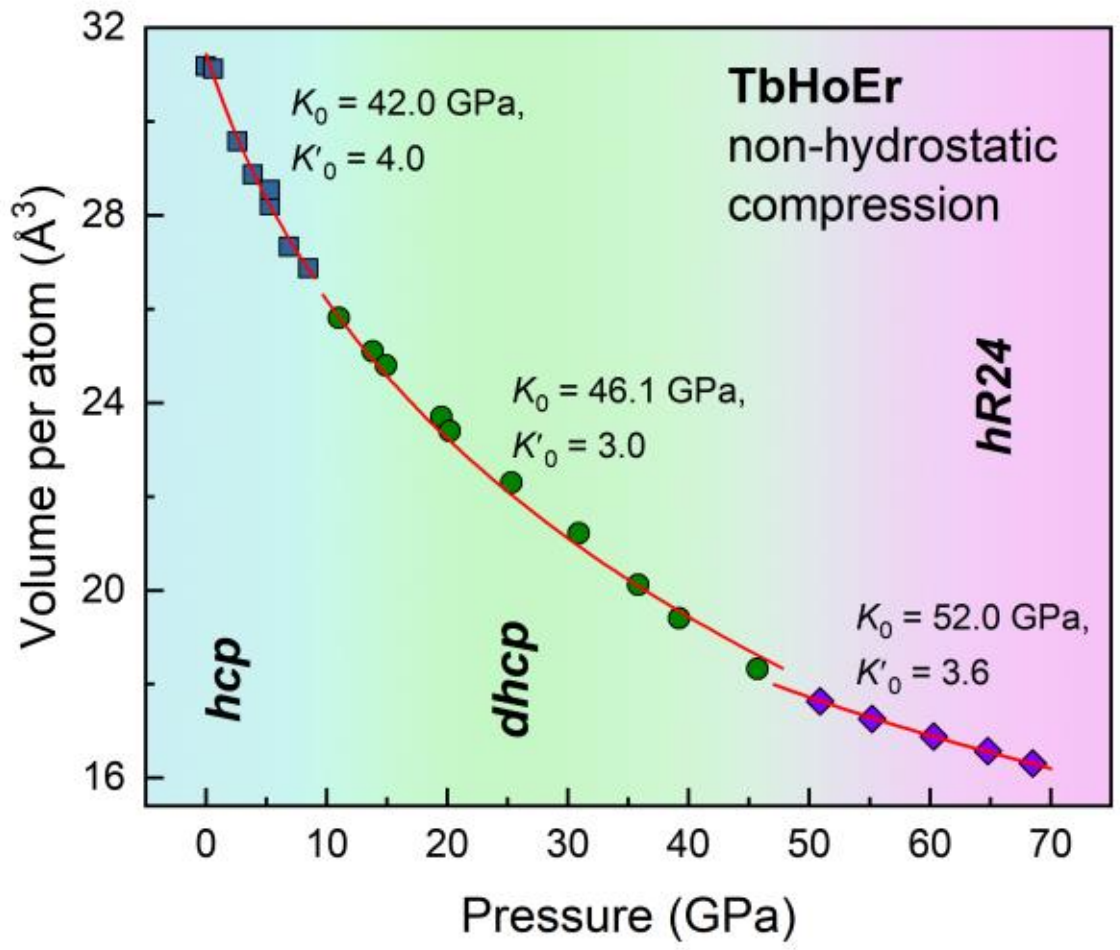

**Figure 8.** Pressure dependence of atomic volume for TbHoEr under non-hydrostatic compression up to 70 GPa. The data show structural evolution from *hcp* to *dhcp* and subsequently to the *hR24* phase. Third-order Birch–Murnaghan EoS fits are shown by red curves for each phase. Colored background regions indicate the approximate pressure ranges over which each phase is observed.

The non-hydrostatic experiment reproduces the initial *hcp*-to-*dhcp* transformation observed under hydrostatic and quasi-hydrostatic conditions, with the transition beginning near 10.5 GPa (see Supplemental Material for details). The major high-pressure reflections are consistent with the *dhcp* indexing used for the neon and silicone-oil datasets. At higher pressure, the extended compression range reveals the emergence of an additional phase beginning near 47–50 GPa. This phase was indexed as a rhombohedral *hR24* structure with space group $R\bar{3}m$ (No. 166). At 60.3 GPa, Le Bail refinement yields lattice parameters of $a$ = $b$ = 5.775(9) Å, and $c$ =

14.024(4) Å in the hexagonal setting (Fig. S4). Thus, under non-hydrostatic compression up to 70 GPa, TbHoEr follows the structural sequence: *hcp* → *dhcp* → *hR24*. The corresponding pressure–volume data and equation-of-state fits are shown in Fig. 8.

The equation-of-state parameters obtained from the non-hydrostatic run are broadly consistent with those derived from the neon and silicone-oil experiments, although modest differences are expected because apparent compressibility in DAC measurements can be influenced by stress state, pressure gradients, and sample-loading geometry. Previous high-pressure studies of multicomponent entropy alloys have shown that pressure-medium conditions, grain size, and deviatoric stress can affect transition behavior and fitted mechanical parameters[33,34]. However, the present TbHoEr results show no drastic pressure-medium dependence of the main structural sequence: hydrostatic, quasi-hydrostatic, and non-hydrostatic experiments all exhibit the initial *hcp*-to-*dhcp* transformation in the same approximate 10–14 GPa pressure interval. Third-order Birch–Murnaghan fits to the non-hydrostatic pressure–volume data yield $K_0 \approx 42.0$ GPa for *hcp*, 46.1 GPa for *dhcp*, and 52.0 GPa for *hR24*, with corresponding pressure derivatives of $K'_0 = 4.0$, 3.0, and 3.6, respectively. These values support the same general compressibility trend observed under pressure-transmitting media, while the differences in fitted parameters are attributed to the expected influence of anisotropic stress and run-to-run experimental conditions.

### C. High-pressure structural evolution of TbHoDy

Synchrotron XRD measurements show that TbHoDy undergoes a pressure-induced *hcp*-to-*dhcp* transformation similar to that observed in TbHoEr. Figure 9 compares selected diffraction patterns from two independent compression runs performed under quasi-hydrostatic conditions using silicone oil as the pressure-transmitting medium. At low pressures, the diffraction peaks are indexed by the ambient *hcp* structure. With increasing pressure, the reflections shift continuously to higher angles, indicating lattice compression. Above approximately 7.5 GPa, additional reflections appear and the low-angle *hcp* peaks begin to split, marking the onset of the transformation to the *dhcp* phase.

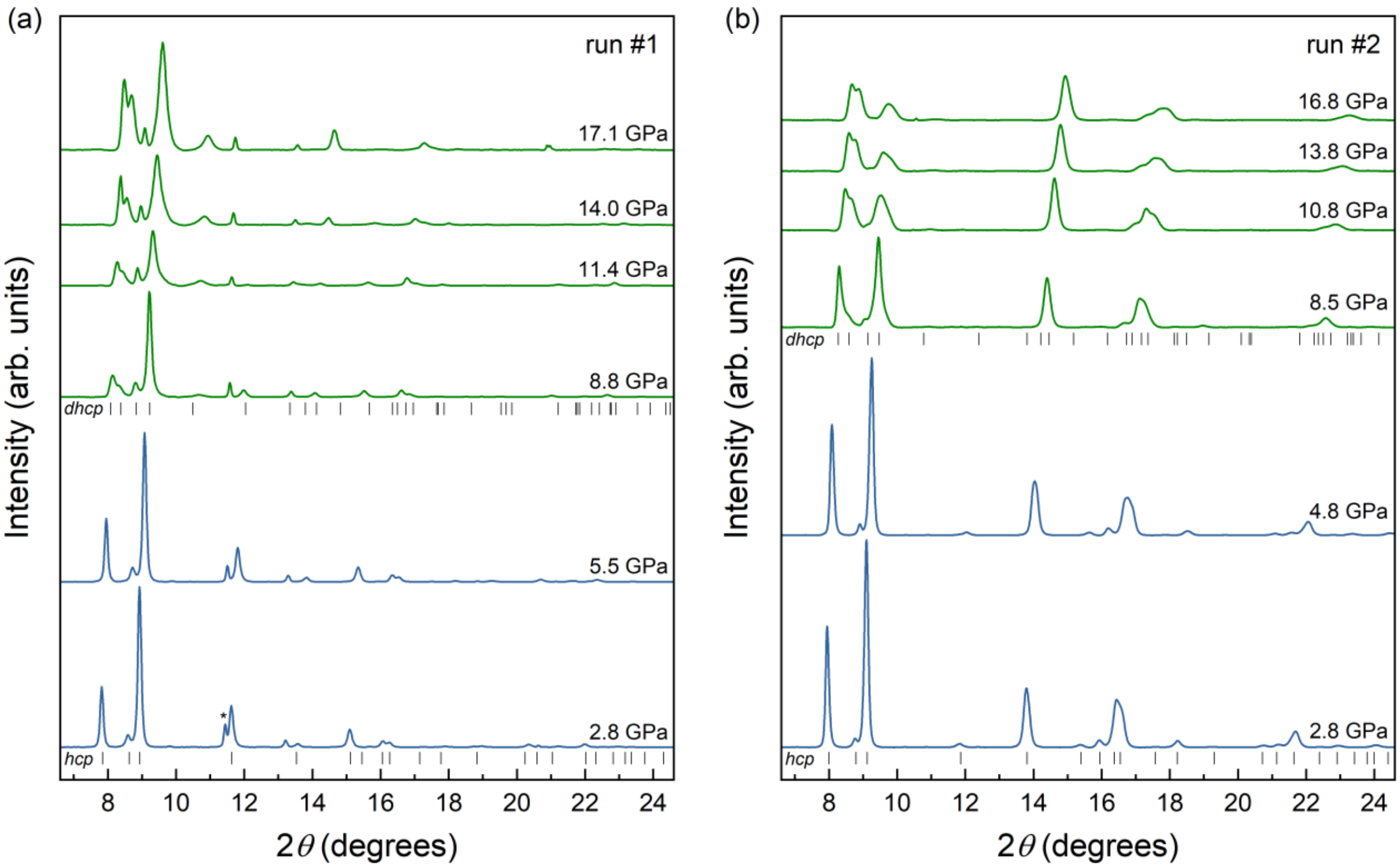

**Figure 9.** Selected synchrotron X-ray diffraction patterns of TbHoDy collected during compression under quasi-hydrostatic conditions using silicone oil as the pressure-transmitting medium: (a) run #1 measured at 16-BM-D using $\lambda$ = 0.4133 Å and (b) run #2 measured at 16-ID-B using $\lambda$ = 0.4246 Å. The diffraction patterns reveal a pressure-induced structural transformation from the ambient *hcp* phase (blue) to the high-pressure *dhcp* phase (green). Vertical tick marks indicate the expected Bragg reflection positions for the corresponding structural models. Background intensity has been subtracted for clarity.

The transition is evident in both runs, although the relative peak intensities and peak shapes differ slightly between the two datasets. These differences are expected for independent DAC loadings and beamline configurations and do not affect the phase assignment. In both cases, the high-pressure diffraction patterns are consistently described by a *dhcp* structure with ABAC stacking. Thus, TbHoDy follows the same primary pressure-induced transformation as TbHoEr, but with the onset shifted to a slightly lower pressure range.

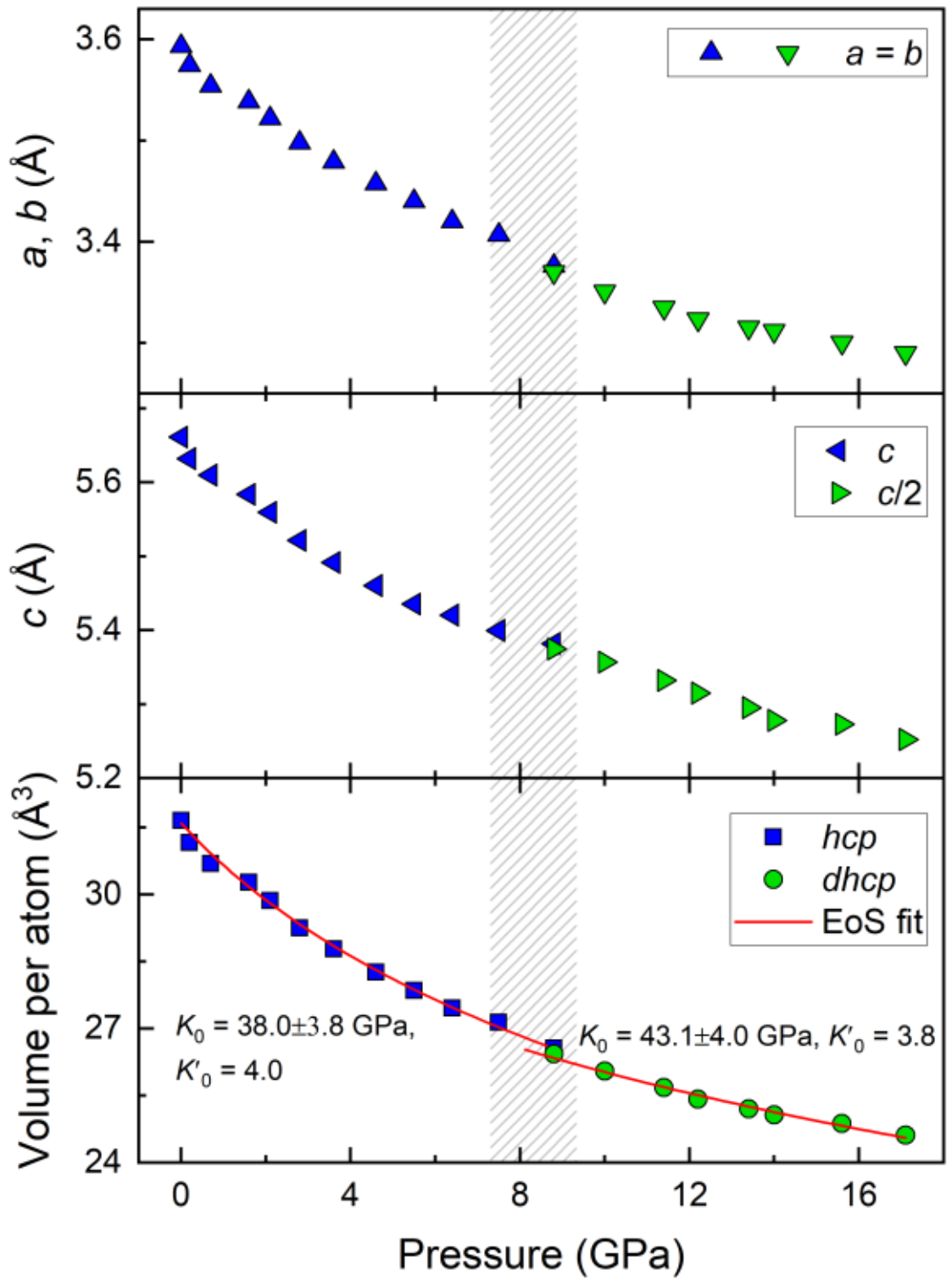

**Figure 10.** Pressure dependence of lattice parameters and atomic volume for TbHoDy extracted from the silicone-oil compression experiment (run #1). Symbols represent experimental data. For the *dhcp* phase, ($c/2$) is plotted to allow direct comparison with the *hcp* ($c$)-axis. The solid red lines are third-order Birch–Murnaghan equation-of-state fits to the pressure–volume data. The shaded region marks the approximate *hcp*-to-*dhcp* transition interval.

Figure 10 shows the pressure dependence of the refined lattice parameters and atomic volume for TbHoDy from the silicone-oil compression experiment. As in TbHoEr, the *dhcp* ($c$)-axis is plotted as ($c/2$) to allow direct comparison with the *hcp* ($c$)-axis. The ($a$)-axis, ($c$)-axis, and atomic volume evolve smoothly through the transition region, supporting interpretation of the *hcp*-to-*dhcp* transformation as a change in close-packed stacking sequence rather than a reconstructive transition with a large volume discontinuity. Third-order Birch–Murnaghan fits to the pressure–volume data yield $K_0 \approx 38.0$ GPa for *hcp*, 43.1 GPa for *dhcp*, with corresponding pressure derivatives fixed to $K'_0 = 4$ and 3.8, respectively. These values indicate that TbHoDy is slightly more compressible than TbHoEr, consistent with the smaller average bulk modulus expected from the constituent rare-earth elements. The increase in $K_0$ across the *hcp*-to-*dhcp* transition also follows the same trend observed in TbHoEr, further supporting a common pressure-induced stacking transformation in these ternary rare-earth MEAs. The slightly lower onset pressure for TbHoDy, approximately 7.5 GPa, compared with TbHoEr, approximately 10 GPa, may be related to its larger ambient unit-cell volume and slightly lower bulk modulus, which arise from replacing Er with the larger Dy atom. This more expanded and more compressible lattice may reach the compression state required to destabilize the hcp stacking arrangement at a lower applied pressure. However, because the transition pressure can also be affected by pressure medium, grain statistics, and sample-loading geometry, this difference should be regarded as a qualitative trend rather than a purely compositional effect.

### D. Suppression of the *Sm-type* phase in medium-entropy rare-earth alloys

The results presented above show that TbHoEr and TbHoDy do not develop a well-resolved bulk *Sm-type* phase during compression. Instead, both alloys undergo a pressure-induced reorganization of close-packed layers that is best described by the development of *dhcp* order. The absence of a systematic set of sharp *Sm-type* reflections is especially significant because the *Sm-type* structure represents a long-period close-packed polytype. Compared with the short ABAB repeat of *hcp* and the four-layer ABAC repeat of *dhcp*, the *Sm-type* structure requires coherent stacking over a 9-layer. Formation of such a long-period motif is therefore more sensitive to local disruptions in stacking registry, strain, and chemical environment.

The two-dimensional XRD images provide direct evidence that the transition region is not structurally simple. As shown in Fig. 11, streak-like diffuse intensity and pronounced azimuthal heterogeneity develop in selected sectors of the diffraction rings for both TbHoEr and TbHoDy. These features appear in the pressure range where the close-packed stacking sequence begins to reorganize. In ideal *hcp*, *dhcp*, or *Sm-type* structures, the stacking periodicity should generate sharp Bragg reflections at well-defined reciprocal-lattice positions. By contrast, faults or interruptions in the stacking sequence redistribute part of the diffracted intensity into diffuse features. Such behavior is well established in diffraction from stacking-faulted close-packed crystals [35], and related streak-like diffuse scattering has been used as evidence of stacking disorder in layered materials such as $\alpha$-$RuCl_3$ [36]. We note that azimuthal intensity depletion in area-detector powder patterns can also arise from incomplete ring statistics due to coarse or textured grains; however, the anisotropy here is concentrated specifically in azimuthal sectors of reflections that carry a *c*-axis component sensitive to stacking along the close-packed direction, and it evolves systematically with pressure in concert with the 1D peak analysis, making a purely grain-statistics origin less likely. It is also noteworthy that the images shown were collected with typical exposure times of 3–5 min per pattern at 16-BM-D, compared with 1–2 s at 16-ID-B; diffuse scattering is spread in reciprocal space rather than concentrated into a narrow Bragg maximum, so its visibility is intrinsically more sensitive to signal-to-noise ratio, and its appearance in the longer-exposure BM-D data is physically unsurprising. Taken together, the diffuse features are most consistently interpreted as evidence of stacking disorder rather than a fully ordered intermediate polytype. This distinction is important: diffuse streaking can arise from disrupted or short-range stacking fragments, but it is not by itself evidence for a well-crystallized *Sm-type* phase. A bulk *Sm-type* phase should produce a systematic set of sharp reflections associated with the 9-layer repeat; since those reflections are not observed, the streaks are more appropriately interpreted as signatures of stacking disorder during the *hcp*-to-*dhcp* transformation.

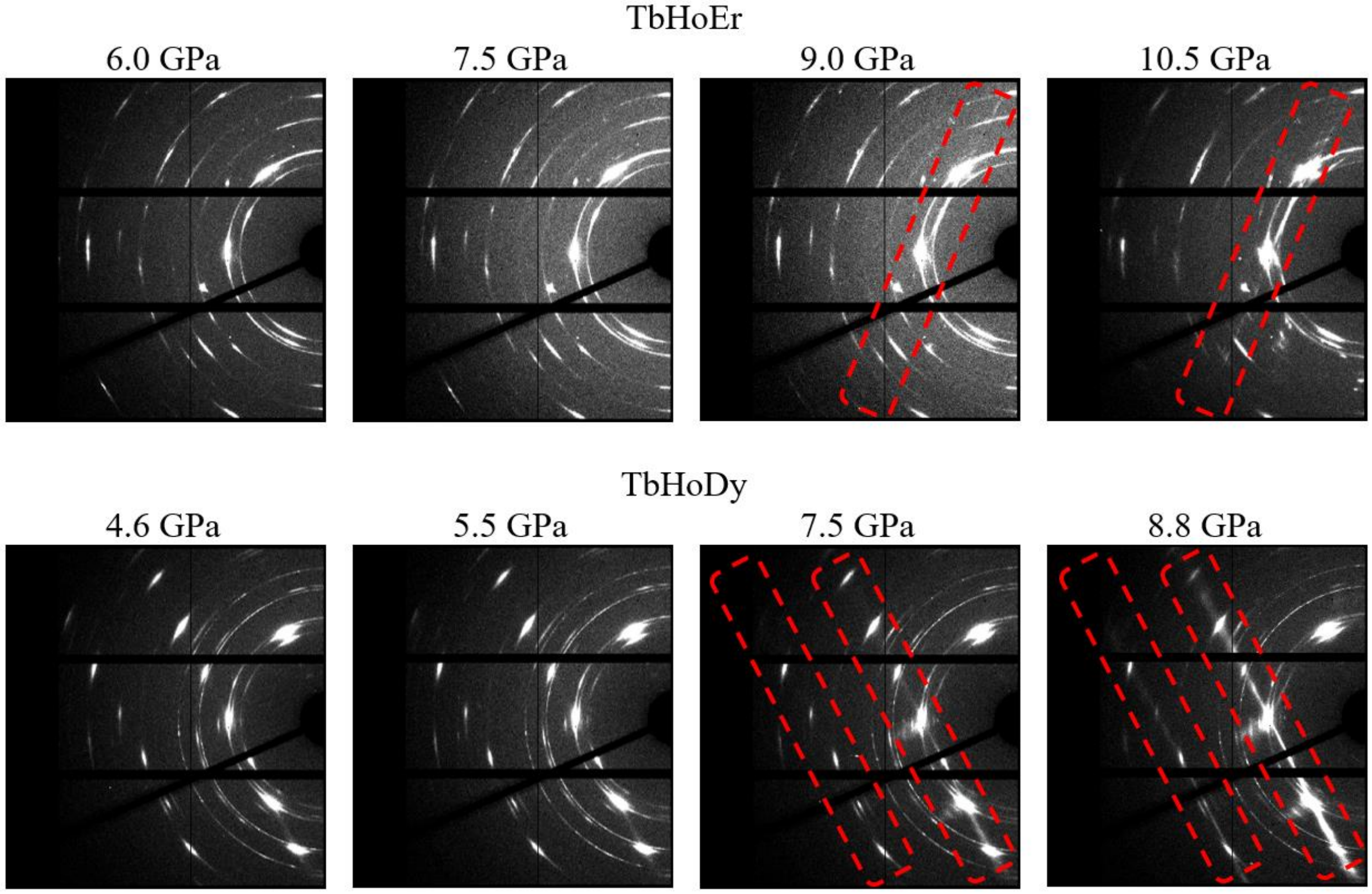

**Figure 11.** Selected 2D synchrotron X-ray diffraction images of TbHoEr and TbHoDy collected during compression at 16-BM-D. Streak-like diffuse intensity develops in selected azimuthal sectors of the diffraction rings, as highlighted by red dashed boxes. The feature is absent or much weaker at lower pressure and is consistent with stacking-disordered close-packed sequences with limited stacking coherence along the close-packed direction. See more details in the Supplemental Material.

To support the qualitative picture of limited stacking coherence, we applied a Scherrer-type analysis [37,38] to the widths of selected diffraction peaks across the pressure series. For a reflection with FWHM $\beta$, measured in radians, at Bragg angle $\theta$, the apparent coherent domain size along the scattering direction is given by the Scherrer relation

$$L_c = \frac{K\lambda}{\beta \cos\theta}, \quad (1)$$

where $K = 0.9$ is a commonly used shape factor and $\lambda$ is the incident X-ray wavelength. Because no instrumental-broadening correction was applied, and because strain, pressure gradients, and texture may also contribute to the measured widths, the extracted values should be interpreted as semi-quantitative lower-bound estimates of apparent coherent domain size rather than rigorous crystallite sizes or stacking-fault densities. Instrumental line widths at 16-BM-D are estimated to be ≤0.03° FWHM, well below the sample peak widths of 0.2°–2.0° observed through the transition. Following the stacking-fault diffraction treatment of Lele [35], the *dhcp* (103) reflection, which has a finite $c$-axis index ($l \neq 0$), is sensitive to disorder in the close-packed stacking sequence, whereas the in-plane (100) reflection ($l = 0$) is comparatively less sensitive to such stacking disorder. We therefore evaluate $L_c$ from the (103) width in the transition region and use the ratio $\beta(103)/\beta(100) > 1$ as a qualitative indicator of stacking anisotropy. In the *hcp* stability fields of both alloys, where the *dhcp* (103) ring is not yet established, this analysis is not applicable. Through the mixed-phase transition region, $L_c(103)$ falls in the range approximately 1–3 nm for TbHoDy (7.5–14.0 GPa) and approximately 1.5–2.3 nm for TbHoEr (10.5–12.1 GPa), while $\beta(103)/\beta(100)$ exceeds unity, reaching values of 1.5–3.7 in the transition region for TbHoDy and 1.6–3.5 for TbHoEr, for both alloys. For comparison, one complete nine-layer *Sm-type* periodicity corresponds to $9 \times (c/6) \approx 9 \times 2.75$ Å $\approx 2.5$ nm using the compressed $c$ lattice parameter. The apparent domain sizes inferred from $\beta(103)$ are therefore comparable to or smaller than a single *Sm-type* repeat length, indicating that the coherence along the stacking direction in the transition region is insufficient to support long-range nine-layer order. These values should be regarded as semi-quantitative supporting evidence, not as rigorous stacking-fault densities; they are consistent with the diffuse features observed in the 2D images and with the absence of sharp *Sm-type* Bragg reflections in the 1D patterns.

Configurational entropy provides one contribution to this behavior, but it should not be treated as the only factor. As mentioned in the introduction, both TbHoEr and TbHoDy have an ideal configurational entropy of $\Delta S_{config} = R\ln3 = 1.10R$, placing them within the medium-entropy regime. This chemical disorder produces a distribution of local environments on the rare-earth sublattice. For a short-period structure such as *hcp* or *dhcp*, this disorder can be accommodated without requiring a unique long-range chemical or stacking arrangement. For a long-period polytype such as the *Sm-type* structure, however, coherent propagation of a specific 9-layer sequence is required over larger distances. Random occupation by different rare-earth atoms can therefore frustrate the development of a single long-period stacking registry, especially when the competing close-packed structures are already close in energy.

Local lattice distortion provides a second contribution. Although Tb, Ho, Er, and Dy are neighboring heavy lanthanides with broadly similar atomic sizes, their elemental *hcp* lattice parameters and atomic volumes are not identical. Using the elemental *hcp* unit-cell volumes, the effective size mismatch in TbHoEr is estimated to be approximately $\delta \approx 0.6\%$, with a maximum effective-radius spread of about 1.5%. The corresponding mismatch for TbHoDy is smaller, approximately $\delta \approx 0.4\%$. These values are modest compared with many transition-metal HEAs, but the relevant energy scale for rare-earth close-packed polytypes is also small. Consequently, even weak local strain fields may bias the system away from a long-period stacking sequence and toward a faulted state or a shorter-repeat *dhcp* structure. This is consistent with the observation that both alloys show the same qualitative *hcp*-to-*dhcp* pathway, while TbHoDy, with smaller size mismatch, appears to show somewhat weaker stacking complexity.

Stacking-fault energetics and transformation kinetics provide a natural framework for understanding this behavior. The *hcp*-to-*dhcp* transformation can occur through changes in the stacking sequence of close-packed planes, without requiring a large reconstructive rearrangement of the metallic framework. By contrast, formation of a coherent *Sm-type* intermediate requires nucleation and propagation of a longer 9-layer stacking repeat. In a chemically disordered MEA, local variations in bond length, magnetic moment, and strain may produce spatially varying stacking-fault energies, favoring finite stacking fragments and faulted close-packed sequences rather than a uniform 9*R* domain.

The comparison with other entropy alloys should therefore be made carefully. Pressure-induced transformations in transition-metal HEAs and MEAs are known to depend on composition, grain size, stress state, and pressure medium[10,33,34]. However, the present rare-earth MEAs differ because their transformation involves competition among close-packed stacking variants rather than a conventional *fcc*-to-*hcp* martensitic transition. Similarly, the observation of a *Sm-type* phase in the quinary rare-earth alloy HoDyYGdTb [27] shows that configurational entropy alone does not universally suppress *Sm-type* order. The present results instead suggest that the structural pathway is governed by the combined balance of configurational disorder, local lattice distortion, stacking-fault energetics, and kinetic accessibility. In this picture, TbHoEr and TbHoDy may transiently sample *hcp*-like, *dhcp*-like, and possibly short 9*R*-like stacking fragments during compression, but these fragments do not evolve into a bulk *Sm-type* phase with long-range crystallographic order.

## 4. Conclusions

In summary, we investigated the high-pressure structural behavior of the medium-entropy rare-earth alloys TbHoEr and TbHoDy using synchrotron X-ray diffraction in diamond anvil cells. Both alloys transform from the ambient *hcp* structure to a high-pressure *dhcp* phase, while no well-resolved bulk *Sm-type* intermediate phase is observed within the studied pressure range. For TbHoEr, extended compression to 70 GPa further reveals the emergence of a high-pressure rhombohedral *hR24* phase. Analysis of two-dimensional diffraction images shows streak-like diffuse scattering in the transition region, providing evidence for stacking disorder and limited coherent stacking domain sizes during compression. These observations indicate that the transformation proceeds through a stacking-disordered close-packed state rather than through a fully ordered *Sm-type* phase.

The suppression of long-range *Sm-type* ordering is proposed to arise from the combined influence of configurational disorder, local lattice distortion, stacking-fault energetics, and transformation kinetics. Together, these factors appear to hinder the development of the coherent 9-layer stacking sequence required for the *Sm-type* structure while favoring the formation of *dhcp* order at higher pressure. More broadly, the present results demonstrate that medium-entropy alloying can modify pressure-induced stacking pathways in heavy rare-earth systems and provide new insight into the interplay between chemical disorder and structural complexity under extreme conditions.

**Declaration of Competing Interest**

The authors declare that they have no known competing financial interests or personal relationships that could have appeared to influence the work reported in this paper.

**Data Availability Statement**

The data that support the findings of this study are available upon reasonable request from the authors.

**Acknowledgements**

This work is supported by the U.S. Department of Energy (DOE) Basic Energy Sciences (BES) Program under award no. DE-SC0023268. This research also used resources of the Advanced Photon Source (DOIs: 10.46936/APS-188276/60013064; 10.46936/APS-189538/60013744; 10.46936/APS-190837/60014665; 10.46936/APS-192702/60016074), a U.S. Department of Energy (DOE) Office of Science User Facility operated for the DOE Office of Science by Argonne National Laboratory under Contract No. DE-AC02-06CH11357. HPCAT operations are supported by DOE–NNSA's Office of Experimental Sciences.